# A charge dependent long-ranged force drives tailored assembly of matter in solution

Sida Wang[†], Rowan Walker-Gibbons[†], Bethany Watkins, Melissa Flynn

& Madhavi Krishnan*

Physical and Theoretical Chemistry Laboratory, Department of Chemistry, University of Oxford, South Parks Road, Oxford OX1 3QZ, United Kingdom

[†] these authors contributed equally

*correspondence to madhavi.krishnan@chem.ox.ac.uk

**Abstract**

The interaction between charged objects in solution is generally expected to recapitulate two central principles of electromagnetics: (i) like-charged objects repel, and (ii) they do so regardless of the sign of their electrical charge. Here we demonstrate experimentally that the solvent plays a hitherto unforeseen but crucial role in interparticle interactions, and importantly, that interactions in the fluid phase can break charge-reversal symmetry. We show that in aqueous solution, negatively charged particles can attract at long range while positively charged particles repel. In solvents that exhibit an inversion of the net molecular dipole at an interface, such as alcohols, we find that the converse can be true: positively charged particles may attract whereas negatives repel. The observations hold across a wide variety of surface chemistries: from inorganic silica and polymeric particles to polyelectrolyte- and polypeptide-coated surfaces in aqueous solution. A theory of interparticle interactions that invokes solvation at an interface explains the observations. Our study establishes a specific and unanticipated mechanism by which the molecular solvent may give rise to a strong and long-ranged force in solution, with immediate ramifications for a variety of particulate and molecular processes including tailored self-assembly, gelation and crystallization, as well as biomolecular condensation, coacervation and phase segregation. These findings also shed light on the solvent-induced interfacial electrical potential - an elusive quantity in electrochemistry and interface science implicated in many natural and technological processes, such as atmospheric chemical reactions, electrochemical energy storage and conversion, and the conduction of ions across cell membranes.

A wealth of natural phenomena emerges from the delicate orchestration of intermolecular and interparticle interactions in the fluid phase. Colloidal particles and biological macromolecules carry ionisable chemical groups which render them electrically charged in solution, and therefore subject to long-ranged electrostatic forces. Our understanding and intuition of molecular behaviour in the fluid phase is grounded in a central principle from classical electromagnetics which dictates that the force between electrical charges of the same sign is not only repulsive at all separations but is symmetric with respect to the sign of the charge. For example, since like-charged objects in vacuum are expected to repel regardless of whether the sign of the charge they carry is positive or negative, the expectation is that like-charged particles in solution must also monotonically repel, particularly at long range where the van der Waals (vdW) attraction is too weak to substantially influence the overall interaction. This view is a hallmark of the Derjaguin-Landau-Verwey-Overbeek (DLVO) theory, a cornerstone of colloid science [1-3]. Nonetheless, clear and well-founded exceptions to the rule exist in the presence of multivalent ions [4,5] and in particular regimes involving high magnitudes of electrical charge and in low dielectric constant media [6]. But over the decades, consistent reports of attraction between like charged particles, nucleic acids, liposomes, polymers and molecular scale matter in aqueous media, particularly in solutions containing low concentrations of monovalent salt - where DLVO theory is expected to hold - have evaded explanation [7-17]. Not surprisingly, this persistent divergence between experiment and theory has received considerable attention in the theoretical literature [6,18-23].

Standing theories invoke a continuum description of the solvent which overlooks finer grained detail such as the molecular nature of the solvent, and its structure and interactions, particularly at an interface with a molecule or a particle [1,2]. Despite the general success of classical mean field theories [3], it is becoming increasingly clear that the molecular nature of water ought to play a defining role in a host of interfacial phenomena in the aqueous phase [24-26]. We recently suggested that the behaviour of the molecular solvent at an interface can give rise to a substantial free energy contribution to the total interaction free energy of two approaching objects carrying electrical charge [27, 28] (Fig. 1a). In particular, for charged matter in water, the model we proposed suggested that the interfacial contribution could be large enough to not only counteract the electrostatic interparticle repulsion but could even overwhelm it, changing the sign of the interparticle force for negatively charged particles and turning it net attractive. The interfacial solvent could therefore be viewed as a pool of free energy that can be tapped by the interparticle interaction, even at long range (Fig. 1a).

At the outset, we briefly summarise the main ingredients of a new model of interparticle interactions in solution (Supplementary section S2.1). We write the total interparticle interaction potential between two particles in solution as the sum of the van der Waals (vdW) contribution, $U_{\text{vdW}}$, the mean field electrostatic contribution, $\Delta F_{\text{el}}$, and an interfacial free energy contribution, $\Delta F_{\text{int}}$, as follows

$$U'_{\text{tot}}(x) = U_{\text{vdW}}(x) + \Delta F_{\text{el}}(x) + \Delta F_{\text{int}}(x) = U_{\text{DLVO}}(x) + \Delta F_{\text{int}}(x) \tag{1}$$

In Eq. 1, $U_{\text{DLVO}} = U_{\text{vdW}} + \Delta F_{\text{el}}$ represents the total interparticle interaction potential as postulated in the DLVO theory.

For large spheres at large distances from each other (say $\kappa x > 2$), the above equation simplifies to

$$U_{\text{tot}}(x) = \Delta F_{\text{el}}(x) + \Delta F_{\text{int}}(x) \approx A\exp(-\kappa_1 x) + B\exp(-\kappa_2 x). \tag{2}$$

Here $A$ and $B$ can be treated as constants, $\kappa^{-1} \approx \frac{0.304}{\sqrt{c_0}}$ nm is the Debye length in an aqueous electrolyte containing monovalent salt at a concentration of $c_0$ moles/litre, and represents the length scale on which electrostatic interactions are screened by counterions in solution [28]. Furthermore, $\kappa_1 \approx \kappa$ is the effective inverse decay length of the electrostatic free energy ($\Delta F_{\text{el}}$) term, and we have $\kappa_1 > \kappa_2 \approx 0.95\kappa_1$ [28]. The interfacial solvation model [28] suggests that

$$B \propto ez\varphi_{\text{int}}\Gamma\frac{d\alpha}{d\psi_{s,\infty}}\psi_{s,\infty}, \tag{3}$$

where

$$\alpha = \frac{1}{1+10^{z(\text{pH}-\text{p}K)}\exp(z\psi_{s,\infty})} \tag{4}$$

is the ionisation probability of a surface ionisable group of valence $z = \pm 1$ in the ionized state [29], which implies

$$z\frac{d\alpha}{d\psi_{s,\infty}} = \frac{-10^{z(\text{pH}-\text{p}K)}\exp(z\psi_{s,\infty})}{[1+10^{z(\text{pH}-\text{p}K)}\exp(z\psi_{s,\infty})]^2} \leq 0. \tag{5}$$

In the above equations, $\Gamma$ denotes the number density of ionisable groups in the particle, while p$H$ and p$K$ denote the negative decadic logarithm of the proton concentration in solution and the equilibrium constant of proton dissociation of the groups respectively. In turn, $\psi_{s,\infty} = \frac{e\phi_{s,\infty}}{k_BT}$ is the dimensionless electrical potential $\phi_{s,\infty}$ created at the surface of an isolated charged object by the ion distribution in the electrolyte, and $e$ is the elementary charge.

Importantly, in Eq. 3, $\varphi_{\text{int}} = \varphi_0 + k\sigma$ is an excess electrical potential at the interface, referenced to the bulk solvent phase, which arises from anisotropic orientation of interfacial solvent molecules, as shown in Fig. 1a and Fig. S8. $\varphi_{\text{int}}$ can be estimated from molecular dynamics (MD) simulations performed for variable surface charge density, $\sigma$, as described previously [27]. The quantity $\varphi_0 = \varphi_{\text{int}}(\sigma = 0)$ is the interfacial electrical potential or solvation potential at a neutral interface in a solvent and is similar to "the potential of zero charge" in electrochemistry and interface science, as well as the air-liquid interfacial potential [30, 31]. For small values of $|\sigma| < 0.1$ $e$/nm$^2$ typical in experiments involving charged particles in solution, we may simplify Eq. 3 by setting $\varphi_{\text{int}} \approx \varphi_0$. Molecular simulation studies provide estimates of $\varphi_0 \approx -0.5$ to $-0.1$ $V < 0$ for the solvation potential an interface immersed in water with respect to the bulk phase [27, 32-36], depending on the molecular model used. Simulations further suggest $\varphi_0 \approx +0.2\ V > 0$ for surfaces immersed in alcohols such as ethanol (EtOH) and isopropanol (IPA) (see supplementary section S1.2).

The main experimental trends suggested by Eqs. 1-5 are: (1) the breaking of charge reversal symmetry in interparticle interactions (e.g., negatively charged particles are characterised by $B < 0$ in water and may therefore attract, whereas positive particles imply $B > 0$, and should therefore repel), (2) use of a solvent in which the sign of $\varphi_0$ is opposite to that of water may reverse this trend (i.e., positive particles may attract whereas negatives repel), and (3) the magnitude of the solvation contribution to the interparticle interaction should depend strongly on the p$H$ in solution. Importantly, when $\frac{d\alpha}{d\psi_{s,\infty}} \to 0$ which occurs outside the in the range of p$H$ given approximately by p$K + 0.2 \lesssim$ p$H \lesssim$ p$K + 2.5$, the electrostatic contribution to the free energy, $\Delta F_{el}(x)$, should be the only surviving term in Eq. 2 (supplementary section S2.1). Here we expect to obtain $U'_{tot} \cong U_{DLVO}$ and the total interaction potential should revert to a form that agrees with the DLVO prediction. We now report on comprehensive experimental tests of the proposed contribution of the interfacial solvent to long range interactions in solution. We show that the interfacial free energy can present a large contribution to the total free energy and can drive the spontaneous formation of ordered assemblies of particles in solution. Furthermore, the sign and magnitude of this contribution have a decisive qualitative impact on the ability of a system of particles to self-assemble, and crucially we show they are amenable to careful tuning using a variety of particle- and solution-dependent parameters.

**Measuring interparticle interactions in a two-dimensional suspension of colloidal particles**

In order to explore the dependence of the interparticle interaction on various system properties such as p$H$ and ionic strength of the electrolyte, particle charge and chemistry, as well as solvent molecular structure, we performed a range of experiments using colloidal particles in solution as interacting entities. We observed the equilibrium spatial structure of a two-dimensional suspension of colloidal particles using bright field optical microscopy (Fig. S3). Analysing the spatial coordinates of all particles in a series of images we constructed radial probability density distributions or pair correlation functions, $g(r)$, that are known reflect the properties of the underlying interparticle interaction potential, $U(x)$. In order to obtain an indication of the form of the underlying interparticle interaction, we performed Brownian Dynamics (BD) simulations with various input potentials of the form in Eq. 2 (supplementary section S1.1). We assumed an initial $U(x)$ profile, generated simulated two-dimensional distributions of particles interacting via this potential, and then determined corresponding simulated $g(r)$ functions (Fig. 1b, top row; Fig. S15-21 and supplementary section S1.1). The parameters $A$ and $B$ in the input potential were then varied to achieve agreement between the measured and simulated $g(r)$s, thus permitting us to identify a pair potential that was qualitatively consistent with the experimental data (Fig. S15-21). We may then compare an approximate pair potential inferred from BD simulations with a total interaction free energy, $U_{\text{tot}}(x)$, calculated for a pair of identical particles using the interfacial solvation model [27] (displayed using open symbols and dashed lines in all figures). As implied by Eqs. 1-5, varying the value of $p = \text{p}K - \text{p}H$ in the theoretical model may result in a calculated $U_{\text{tot}}(x)$ that is purely repulsive, purely attractive, or a non-monotonic function displaying a minimum of depth $w$ at an intersurface separation $x_{\min}$. Parameter values for all calculated curves in this study are quoted in Table S1 (see supplementary information). Note that our BD simulations assume that all interactions between particles are pairwise additive, which could be a reasonable approximation at large interparticle separations under our experimental conditions, but may not be rigorously true at all separations [37].

**Negative particles attract and positive particles repel in water**

We first examined a system of colloidal silica particles of nominal diameter 4.82 μm dispersed in deionised water with an estimated background ionic strength of $c_0 =$ 5 μM. The measured electrical potential (zeta potential) in the vicinity of the particle surface, $\zeta \approx -40$ mV, is indicative of a strongly negative surface charge arising from a high density of ionized surface silanol groups. DLVO theory predicts strong interparticle repulsions under these conditions implying the observation of a randomly dispersed 2D distribution of particles. In marked departure from the intuitive expectation, we noted that the particles spontaneously self-assembled into stable, slowly reorganising, hexagonally close packed (hcp) clusters characterised by an intersurface separation on the order of $x \sim 0.5$ μm $\approx 5\kappa^{-1}$, similar to previous experimental reports [12, 38, 39] (Fig. 1b (middle), Supplementary Video 1). The spatial structure of the clusters, and the presence of large voids between clusters, implies the presence of a strong, long-ranged attractive interaction between particles of like charge, which is counteracted by a repulsion at shorter range, giving rise to a stable minimum at comparatively long range [13]. The measured $g(r)$ profiles present periodic peaks of diminishing height reflecting the ordered internal structure of finite-sized clusters - a characteristic signature of attractive interparticle interactions in solution (Fig. 1c, top panel). BD simulations (snapshots in Fig. 1b, top row) permit us to extract approximate pair potentials, $U(x)$, characterising the underlying interaction (Fig. 1d, solid curves). The measured $U(x)$ profiles agree with calculations from the interfacial solvation model which capture both the approximate location of the minima, $x_{\min}$, as well as their depth, $w$ (Fig. 1d, dashed lines with symbols). Experiments in the range of ionic strengths from 5 μM - 1 mM on negatively charged silica and carboxylated particles–show that an attractive interfacial contribution to the total free energy can produce minima of comparable depth, $w$, in the interparticle potential at intersurface separations $x_{\min} \approx 2 - 10\kappa^{-1}$, as suggested by the interfacial solvation model [27,] [28]. Finally, interaction potentials measured for isolated pairs of interacting particles display good agreement with the $w$ and $x_{\min}$ values inferred from the $g(r)$ measurements (see Fig. S5 and supplementary methods). These direct pair-potential measurements confirm that the observed cluster formation stems from a pair interaction potential that indeed carries a substantial attractive component, rather than from collective interactions in a system of purely repulsive particles [20].

Turning our attention to the behaviour of positively charged aminated silica particles under the same experimental conditions, we found that particles remained randomly dispersed in solution as expected for a purely repulsive interparticle interaction (Fig. 1b, lowest row, Supplementary Video 1). The measured $g(r)$ profiles were relatively featureless, and categorically devoid of the periodic structure characteristic of ordered clusters (Fig. 1c, lower panel). The corresponding $U(x)$ profiles imply monotonically decaying repulsions (Fig. 1d, pink solid lines) in good agreement with the profiles calculated from the interfacial solvation model (Fig. 1d, pink dashed lines). The conspicuous absence of a long-ranged minimum in the pair interaction potential for positive particles constitutes a successful preliminary test of a key characteristic of the interfacial solvation model, namely that the total interparticle interaction in solution can be either repulsive or attractive depending on the sign of charge on the particle. In other words, the solvent appears to be responsible for this profound qualitative departure from charge reversal symmetry in interparticle interactions.

**p$H$ controls the magnitude of the long-ranged force**

Next we explored the influence of p$H$ on the interaction between silica particles. Silica surfaces carry a range of different ionisable silanol groups that are characterised by different number densities and widely different p$K$ values between 2 and 11 [40, 41]. A range of different p$K$ values implies a wide range p$H$ values over which the surface ionisable groups may undergo a change in their extent of ionisation [29, 41]. We performed particle interaction experiments at a fixed ionic strength but varied the p$H$ in solution from 4 to 10 and examined the experimentally inferred value of $w$. Note that $w < 0$ denotes the minimum value of the free energy in a measured pair-potential and we take $w = 0$ for a purely repulsive measured pair-potential. For silica, we found that although the magnitude of $w$ decreased significantly with increasing p$H$ it remained substantial ( $|w| \approx 2\ k_\mathrm{B}T$) even at high p$H > 9$ (Fig. 2d, Supplementary Video 2). This occurs despite the fact that the electrostatic interaction, $\Delta F_\mathrm{el}$, is expected to grow in magnitude with increasing p$H$, progressively counteracting an attractive interfacial contribution that we expect to simultaneously weaken since $\frac{\mathrm{d}\alpha}{\mathrm{d}\psi_\mathrm{s}} \to 0$. Interestingly, we found that the observations can be qualitatively explained within the interfacial solvation model incorporating a broad distribution in p$K$ values known to describe silica surfaces (Fig. 2a, d top panels; see supplementary section S2.5, Supplementary Video 2).

In order to perform a more quantitative exploration of the effect of p$H$, we examined the behaviour of carboxylated-melamine resin (COOH) particles. Here we assume that the charge on the particle is conferred by ionization of a single chemical species of p$K \approx 4.5$, the nominal value for an isolated carboxyl group. We observed a clear maximum in the depth of the potential minimum, $|w|$, in the range of p$K <$ p$H <$ p$K + 3$, in qualitative agreement with the trend expected for $\frac{\mathrm{d}\alpha}{\mathrm{d}\psi_\mathrm{s}}$, as given by Eq. 3 and discussed further in the supplementary information. (Fig. 2a, d middle panels, supplementary section S2.5, Supplementary Video 3).

Next, we examined the behaviour of positively charged aminated silica particles carrying a high density of amino groups ($NH_2$) of nominal p$K \approx 9.5$. We did not observe cluster formation in positive particles over the entire range of p$H$ (Fig. 2a, bottom panel; Supplementary Video 4). In fact the $g(r)$ curves reflected robust interparticle repulsions at large distances that were strongest at low values of p$H \approx 4$ (Fig. 2b, bottom panel). Under more alkaline conditions given by 7< p$H <$ p$K \approx 9$, the net charge on the particles decreases due to gradual deprotonation of the amino groups and down regulation of particle charge.

Discharging of the amino groups correlates with a significant reduction in measured zeta potential and a consequent weakening of the interparticle repulsion (Fig. 2b, bottom panel), but the formation of stable, ordered clusters was never observed (Fig. 2a-c). This range of p$H$ values (7< p$H$ < 9) may be contrasted with the equivalent p$H$ range of 5-7 for carboxyl particles where we in fact observed the most stable clusters characterised by the deepest minima in interaction energy (largest $|w|$) generated by the long-ranged attractive force. We obtained similar results for silica particles coated with positively charged polyelectrolytes such as polyethyleneimine (PEI) (Fig. S13). Thus we find that the measured p$H$ dependence of interparticle interactions - for both signs of particle charge and a range of surface chemistries - carries key signatures of the solvation free energy contribution, furnishing crucial validation of an interaction mechanism invoking charge dependent solvent structure at the solid-liquid interface.

**The sign of charge on the particle governs the sign of the long-ranged force**

In order to further probe the influence of surface chemistry on the interparticle interaction we examined the behaviour of negatively charged polystyrene sulfonate (PSS) surfaces. Isolated styrene sulfonic acid groups are highly acidic in free solution (p$K < -0.5$) but are expected to be characterised by p$K \approx 3$ in the context of a polymer backbone [42]. We produced PSS coated silica particles using the layer-by-layer polyelectrolyte film deposition method [43]. We first coated silica particles with a layer of positively charged poly(diallyldimethylammonium chloride) (PDADMAC) polymer, which switched the sign of the zeta potential from $\zeta \approx -50$ mV to $\zeta \approx +40$ mV. PDADMAC coated silica particles in solution no longer formed clusters (Fig. 3). We then further coated PDADMAC-coated silica particles with a layer of PSS which altered the zeta potential from $\approx +40$ mV to $-70$ mV. We found that the system of PSS-coated particles formed stable clusters in solution at p$H$ 5-6 ($\approx$ p$K + 2.5$), indicating the reappearance of long range interparticle attractions, similar to uncoated silica particles. We then proceeded to sequentially coat these PSS particles with alternating layers of PDADMAC and PSS, confirming the sign of the particle charge after each coating procedure using zeta potential measurements. We consistently found that the long ranged interparticle attraction could be switched on and off depending on the sign of charge of the most recent surface coating. Data from 4 layers of sequential coating are presented in Fig. 3 (Supplementary Video 5). Similar results were obtained with alternating coatings of positively charged PEI and negatively charged PSS (Fig. S20). All polyelectrolyte coating experiments were performed at $c_0 = 15$ μM (Fig. 3 and Fig. S20).

To continue exploring the generality of the 'like-charge assembly' phenomenon we turned our attention to the examination of the interactions of polypeptide surface coatings on silica particles. Polylysine (poly-K) and polyglutamic acid (poly-E) polypeptides are routinely used as surface coatings, generating positively and negatively charged surfaces on account of their ionized amino ($NH_3^{+}$; $pK \approx 10$) and carboxyl ($COO^-$; $pK \approx 4.5$) sidechain groups respectively. Here again we found that whereas positively charged lysine surfaces repelled as expected (similar to PEI and PDADMAC-coated particles), poly-E surfaces displayed long range attractions similar to our carboxylated melamine resin particles (Fig. 3d). Once again we were able to repeatedly alternate between attractions and repulsions in the particle suspension by sequentially coating the particles with poly-K and poly-E layers. Beyond cementing the generality of the like charge attraction effect, the observations for polypeptide surface coatings have important implications for archetypal notions in biomolecular interactions. In particular these findings may call for a re-evaluation of the general expectation that negatively charged poly-E stretches in proteins repel unequivocally in solution. Our work suggests that this does not necessarily hold when the p$H$ in solution approaches a value that is within about 2-3 units higher than the p$K$ of the ionisable groups (Fig. 2, middle panels), and that like-charged molecules in solution may in fact experience a counter-intuitive strong and long-ranged attraction, even under physiological conditions, as discussed later.

The above experiments show that negatively charged silica surfaces ($pK \approx 2 - 11$), carboxylated melamine resin particles ($pK \approx 4.5$), polystyrene sulfonate ($pK \approx 3$), and polyglutamate ($pK \approx 4.5$) coated polypeptide surfaces suspended in water support long range interparticle attractions when the $pH$ of the solution lies within a range of about 3 units above the $pK$ of the ionisable groups. In contrast, positively charged particles display robust repulsions regardless of surface chemistry, and no evidence of a long-ranged attraction even when $pH \approx pK$.

It is worth noting here that although our theoretical results for interfacial solvation invoke molecular simulation data for model oxygen-atom surfaces, recent MD simulation studies on water orientation at a variety of more realistic surfaces whose chemistry ranges from that of inorganic silica to polymeric surfaces clearly reveal that essential features of interfacial water orientation and the sign of the associated interfacial potential, $\varphi_0$, are preserved across these materials [32]. It is possible that the dominant role played by hydrogen bonding in water is responsible for the relative independence of qualitative aspects of the orientational behaviour of water at an interface on specific surface chemistry [32-34]. Interfaces with organic solvents may however be expected to show significant dependence of both sign and magnitude of $\varphi_0$ on surface chemistry [32].

**Chemically dissimilar particles form ordered clusters**

In order to illustrate the fundamental dependence of the long-ranged attraction on electrical charge and the charged state of the ionisable groups we performed experiments involving mixtures of particles of different surface chemistry and observed the nature of the crystallites formed. First we performed experiments on mixtures of COOH particles and $SiO_2$ particles at high values of $pH$ - much larger than the typical $pK$ value of either species. Cluster formation was neither observed in particle mixtures nor for the pure species, as expected. Next, at $pH$ values where each species is known to form clusters in the pure state (Fig. 2), we found that the mixture formed hcp clusters composed of both species of particle with signatures in the $g(r)$ reflecting both particle sizes. We then performed experiments at $pH \approx 4$ where pure COOH particles do not form clusters, but SiO$_2$ particles do cluster. Interestingly, we found that rather than exclude the non-cluster forming COOH species, the system displayed stable hcp clusters containing mixtures of both types of particles (see Supplementary Video 6). All these observations can be explained within the proposed interfacial solvation model of interparticle interactions (see Supplementary section S2.6). Specifically, it appears sufficient for one species of particle to be in the charge regulating regime ($\frac{d\alpha}{d\psi_s} \neq 0$) in order for an attraction between chemically dissimilar like-charged particles to manifest. We thus show that similar or identical chemistry is not a requirement for interparticle attraction, but that the electrical charge-state of the groups underpins a force with a general and tunable character that holds in a broad sense. Simply put, chemical properties of the particle enter the physics of the interaction via the p$K$ of the ionisable groups. The observations could carry implications for the biological phase separation problem e.g., where strongly acidic RNA molecules, that are not expected to cluster on their own under physiological conditions, do however participate in cluster formation in mixtures involving other charged molecular species such as proteins.

Finally we point out that although our simple model of the interaction between a pair of identical particles captures the qualitative features of the experimental observations, this framework does not quantitatively capture some aspects of the experimentally observed interaction. In particular, for experiments on negatively charged particles at ionic strengths $c_0 \gtrsim 0.1$ mM, the present model of the total pairwise interparticle interaction over-estimates the magnitude of the attractive interfacial contribution, $\Delta F_{\text{int}}$, at smaller separations, $x < 200$ nm (Fig. 2c, middle panel). Thus, in the range of $p$ values where the interfacial attraction is dominant ($\frac{d\alpha}{d\psi_s}$ is large), our current model often suggests that the overall interaction is net attractive, i.e., $U'_{\text{tot}}(x)$ is monotonically decreasing with decreasing separation, whereas experiments display a robust net repulsion at shorter range ($x < 200$ nm) in conjunction with a long-ranged attractive interaction (Fig. 2c middle panel). In future, a more comprehensive theoretical description of interfacial solvation in the pair interaction may furnish better quantitative agreement between experiments and theory (see supplementary section S2.4).

**Negative self-attraction in water switches to positive self-attraction in alcohols**

The question then naturally arises as to whether the type of charge-asymmetry observed in the interparticle interaction in water could be altered in a different solvent on account of differences in molecular orientational behaviour at an interface. In ethanol (EtOH) and isopropanol (IPA) for example, molecular simulations show that the average molecular dipole is inverted with respect to that of molecular water at a neutral surface, i.e., interfacial molecules orient such that their comparatively hydrophobic methyl (-CH$_3$) groups point towards a neutral surface (Fig. 5a, Fig. S8b; see supplementary section S1.2) [44]. This molecular orientation gives an interfacial solvation potential at zero surface charge that has a positive sign, i.e., $\varphi_0 > 0$, which may be contrasted with water where the electronegative O-atom orients towards a neutral surface on average, yielding $\varphi_0 < 0$. Previous attempts to probe the interfacial electrical potential have suggested similar orientational behaviour and an inversion in sign of $\varphi_0$ for alcohols at interfaces compared to water [30, 45, 46]. As indicated in Eq. 3, our model of interfacial solvation would suggest that positively charged particles suspended in these alcohols should attract at long range, in contrast with observations on the same type of particle in water.

We found that positively charged amine-derivatized silica particles suspended in EtOH and IPA displayed positive zeta potentials (Fig. 5c). We recall that these particles remained well dispersed when suspended in water (Fig. 1 and Fig. 2). In EtOH and IPA however, we found that positively charged particles formed strong and stable clusters, displaying $g(r)$ distributions reminiscent of attractive interactions generally encountered for negatively charged silica particles in water (Supplementary Video 7). Negatively charged carboxylated particles on the other hand remained well dispersed in both alcohols, did not form clusters, and qualitatively echoed observations on positively charged particles in water. The experimental observations are qualitatively captured by calculations of $U_{\text{tot}}(x)$ as shown in Fig. 5d, yielding physically plausible model parameters (see Supplementary Video 7, supplementary section S2.2, and Table S1).

**A role for interfacial solvation in biological phase segregation**

Finally, inspired by the spontaneous formation of tuneable clusters in like-charged microspheres in solution, we examine the implications of this long ranged solvation force for the formation of reversible condensates at the molecular scale under physiological conditions, broadly encountered e.g., in the liquid-liquid phase segregation problem in biology. We constructed a simple model of a protein as a sphere carrying surface ionisable groups exposed to the solvent (Fig. 6a). We calculated the pair interaction energy of two spherical, charged molecules carrying 5 ionisable glutamic acid groups ($pK = 4.5$), immersed in an electrolyte containing monovalent salt at a concentration in the range $c_0 = 10$ mM-1 M in the pH range 5.5-8 (Fig. 6, Supplementary Fig. S11). We then examined the sign and magnitude of the total interaction potential, $U'_{\text{tot}}(x)$, given by Eq. 1, at a nominal intersurface distance of $x = 2$ nm between the particles. First, we note that the interparticle electrostatic repulsion, $\Delta F_{\text{el}}$, is expected to be significant over the range of $pH$ and salt concentration considered since the molecules carry substantial net charge (of the same sign) (Fig. 6c, d). However, upon inclusion of the interfacial contribution to the total free energy we found that the interaction could be expected to turn significantly attractive (i.e., $U_{\text{tot}}(x) < 0$), particularly at lower ionic strength and at $pH$ values significantly higher than the $pK$ of the charged groups (Fig. S11, see supplementary section S2.7). Estimating the profile of a coexistence curve demarcating parts of the parameter space where phase separation may occur from regions where the molecules remain well dispersed, we find that the interfacial model suggests the formation of clusters or droplets in solution at salt concentrations up to $c_0 \approx 0.5$ M and at p$H$ values as high as $\approx 6.5$ (Fig. 6d, right panel; see supplementary section S2.7). In fact, the p$H$ and salt concentration dependent phase separation behaviour suggested by our model is qualitatively similar to experimental observations reported, e.g., for the negatively charged wild-type yeast protein Sup35 that carries a net excess of 6 acidic amino acids (aspartic and glutamic acids) over the number of basic side chains (lysine and arginine) in the domain that controls its phase segregation response (Fig. 6b) [47]. A DLVO-type model of the same interaction however would not capture the experimental trends reported for Sup35 (Fig. 6d, left panel; see supplementary section S2.7). Future work could examine the use of molecular simulations that incorporate both explicit water and charge regulation of ionisable protein groups in order to construct more accurate models of interaction free energies for a variety of molecular systems [48]. Note that the charge-asymmetry of the interfacial solvation contribution we suggest by no means precludes coacervate formation in positively charged molecular systems containing aromatic rings. Here, condensate formation is believed to be driven by an attractive cation-π interaction [49].

**Conclusion**

In conclusion, the ability to *qualitatively* alter the attractive or repulsive nature of an interparticle interaction at long range by tailoring the surface properties of the particle points to an interfacially governed contribution to the total interaction free energy of a pair of particles in solution. The ability to tune the interaction using the solution p$H$ and particle p$K$ strongly suggests that charge regulation plays a defining role in the magnitude of the free energy contribution. The interparticle interaction further displays a stark qualitative asymmetry with respect to the charge on the particle such that *the sign of the particle charge determines the sign of the force* at long range. Importantly, the type of charge-reversal asymmetry observed in water can be inverted through the use of a solvent with different structural and interaction properties. Our proposed mechanism invoking the excess solvation potential at an interface captures all these features and explains the experimental observations. Overall, we provide compelling evidence for a profoundly different view of interparticle and intermolecular interactions in the fluid phase. We show that the structural and chemical asymmetry inherent to the solvent molecule, and the resulting molecular orientational anisotropy at an interface, can turn the interaction between like-charged objects attractive, driving the formation of stable, ordered clusters and assemblies. As a rule of thumb, when the sign of $\varphi_0$ matches that of the sign of charge of the particle, attraction may be expected between like charged particles. The remit of the problem specifically covers p$H$ and salt-concentration dependent cluster or phase formation, as reported in a range of chemical and biochemical processes. Examples of relevant assembly processes range from gelation, coagulation, crystallisation and biological phase segregation to histone-modulated packaging of DNA, virus binding to sulfonated carbohydrates on host cells, biofouling, or indeed any experimental situation where counterintuitive attractions are implicated between suspended entities carrying net electrical charge of the same sign (negative in the aqueous phase). Our study thus postulates a specific, molecular charge-dependent role for water in driving self-assembly and clustering of molecules and particles, relevant even, e.g., to the origin of life.

Our results have further ramifications for interactions in solvent mixtures, and in solutions containing additives such as zwitterions, osmolytes, polyols or other overall electrically neutral molecules that are not expected to strongly influence the electrostatic interaction *per se* [50]. The solvation structure at an interface, which gives rise to the solvation potential, is likely to be influenced by the composition of the electrolyte. Net-neutral molecular species in solution may indeed alter the solvation potential arising from the additive-free background electrolyte and thus modify the total interaction between charged particles. In summary, we suggest that as a general principle, the concerted action of electrical charge at an interface and a local non-zero solvation potential - either or both of which may vary with separation - can give rise to substantial contributions to the interaction free energy between entities in suspension, with the sign and magnitude of the solvation potential influencing those of the resulting interparticle force. The concept invoked here for *inter*particle and *inter*molecular interactions may well be found to hold for *intra*molecular interactions and conformational changes associated with biomolecular folding. Finally, our findings provide evidence for the ability to probe the sign and magnitude of the interfacial electrical potential due to the solvent, which has so far been believed to be immeasurable [30, 45, 51].

**Figures**

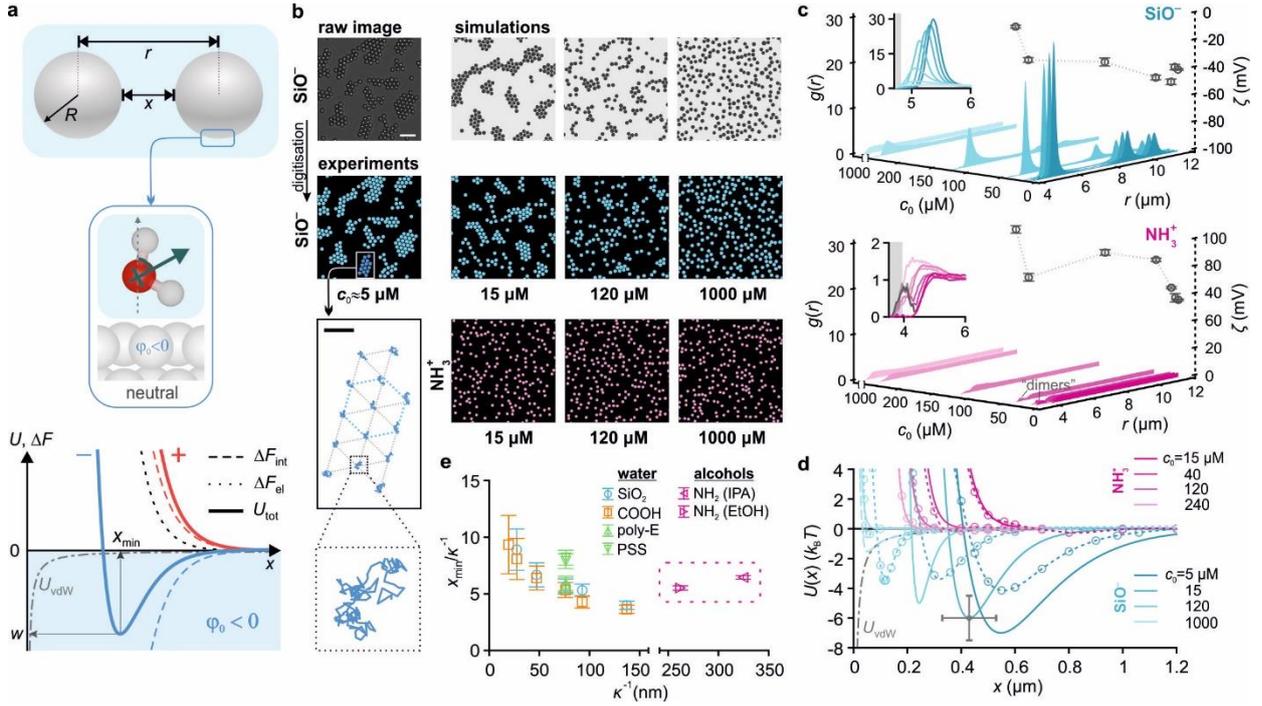

**Fig. 1. Interparticle interactions in solution can break charge reversal symmetry.** (**a**) Schematic representation of two interacting particles as spheres of radius $R$, at an interparticle separation $r$ and intersurface separation $x$ (top). Depiction of the average orientation of water molecules closest to a neutral particle surface carrying no electrical charge (middle). The electronegative oxygen atom of water points towards the interface whereas the hydrogen atoms point slightly towards the bulk, giving rise to an interfacial or solvation potential $\varphi_0 < 0$ which does not arise in a smooth-continuum description of water. The molecular dipole moment (green arrow) points away from the surface. Fig. S8a presents the variation in interfacial potential, $\varphi_{\text{int}}$, and the corresponding interfacial free energy, $f_{\text{int}}$, as a function of surface charge density, $\sigma$, obtained from MD simulations. Bottom panel: Various contributions to the total free energy of interaction for the interaction of positively (red lines) and negatively charged (blue lines) particles: electrostatic repulsion, $\Delta F_{\text{el}}$ (black dotted line and applies to both signs of charge), interfacial solvation contribution, $\Delta F_{\text{int}}$, which depends on the sign of charge (dashed line), total interaction energy, $U_{\text{tot}} = \Delta F_{\text{el}} + \Delta F_{\text{int}}$ (solid lines), and the small vdW contribution, $U_{\text{vdW}}$ (dashed-dotted grey line, applies to both cases). When $\varphi_0 < 0$, negatively charged particles in solution are expected to display a minimum in $U_{\text{tot}}$ of depth $w$ at an interparticle separation $x_{\text{min}}$ (blue solid line), whereas positively charged particles are expected to repel monotonically (red solid line), displaying no minimum. (**b**) Negatively charged silica particles of radius $R = 4.82$ μm (SiO⁻) in water (scale bar: 20 μm, light blue particles) form hexagonally close packed (hcp) clusters which decrease in size and vanish altogether with increase salt concentration, $c_0$ (middle row). Inset (bottom left) presents measured spatial trajectories of silica particles in a single cluster over a period of 30s, and highlights the hcp structure of the clusters (scale bar: 5 μm). Positively charged aminated silica particles (NH₂, pink particles) of radius $R = 3.92$ μm do not form clusters in water regardless of $c_0$ (bottom row). Images are recorded using bright field microscopic observation of particles as presented in Fig. S3 and Fig. S22 and are digitised and false-coloured to facilitate presentation.BD simulation snapshots for interactions of negatively charged particles (top row). (**c**) Measured radial probability density function $g(r)$s for negatively charged (top) and positively charged particles (bottom), including measured zeta potentials ($\zeta$) (circular symbols) for various values of $c_0$. Multiple periodic peaks in the $g(r)$ reflect the formation of ordered particle clusters (top). The small peak around $2R$ in some cases, reflects a small fraction of particle "dimers" in the sample (bottom inset). Shaded region in insets indicates $r \leq 2R$. (**d**) Pair interaction potentials of the form $U(x) = A\exp(-\kappa_1 x) + B\exp(-\kappa_2 x)$ inferred for the experimental data from BD simulations (solid lines) for negative particles (blue) and positive particles (pink). Calculated $U_{\text{tot}}(x)$ curves as described in Ref. 27 and supplementary sections S2.1 and 3.2. (dashed lines with symbols; see Supplementary Table S1 for parameter values). At $c_0 = 1$ mM, the potential inferred from BD simulations for negatively charged particles is of the form $U(x) = A\exp(-\kappa_1 x) + U_{\text{vdW}}$ (light blue dashed line) indicating that the contribution from the $\Delta F_{\text{int}}$ term has effectively vanished, with the solid blue line representing the repulsive term in the interaction alone (see Supplementary Figs. S15-16 and Tables

S2-3 for details). (**e**) Location of the minimum, $x_{\min}$, as a function of Debye length, $\kappa^{-1}$, shows that minima in $U(x)$ generally appear in the range $x_{\min} \approx 5 - 10\kappa^{-1}$ for negatively charged particles – silica ($SiO_2$, light blue symbols), carboxylated melamine resin (COOH, orange symbols), polyglutamic acid (poly-E, green symbols) and polystyrene sulfonate (PSS, green symbols) - in water, and positively charged particles ($NH_2$, pink symbols) in alcohols (see Fig. 5). Error bars depict uncertainties in $x_{\min}$ values and $U(x)$ curves arise both from a nominal $\pm 100$ nm uncertainty on particle diameters and an estimated 1 $k_\mathrm{B}T$ uncertainty in matching experimental and simulated $g(r)$ curves, respectively.

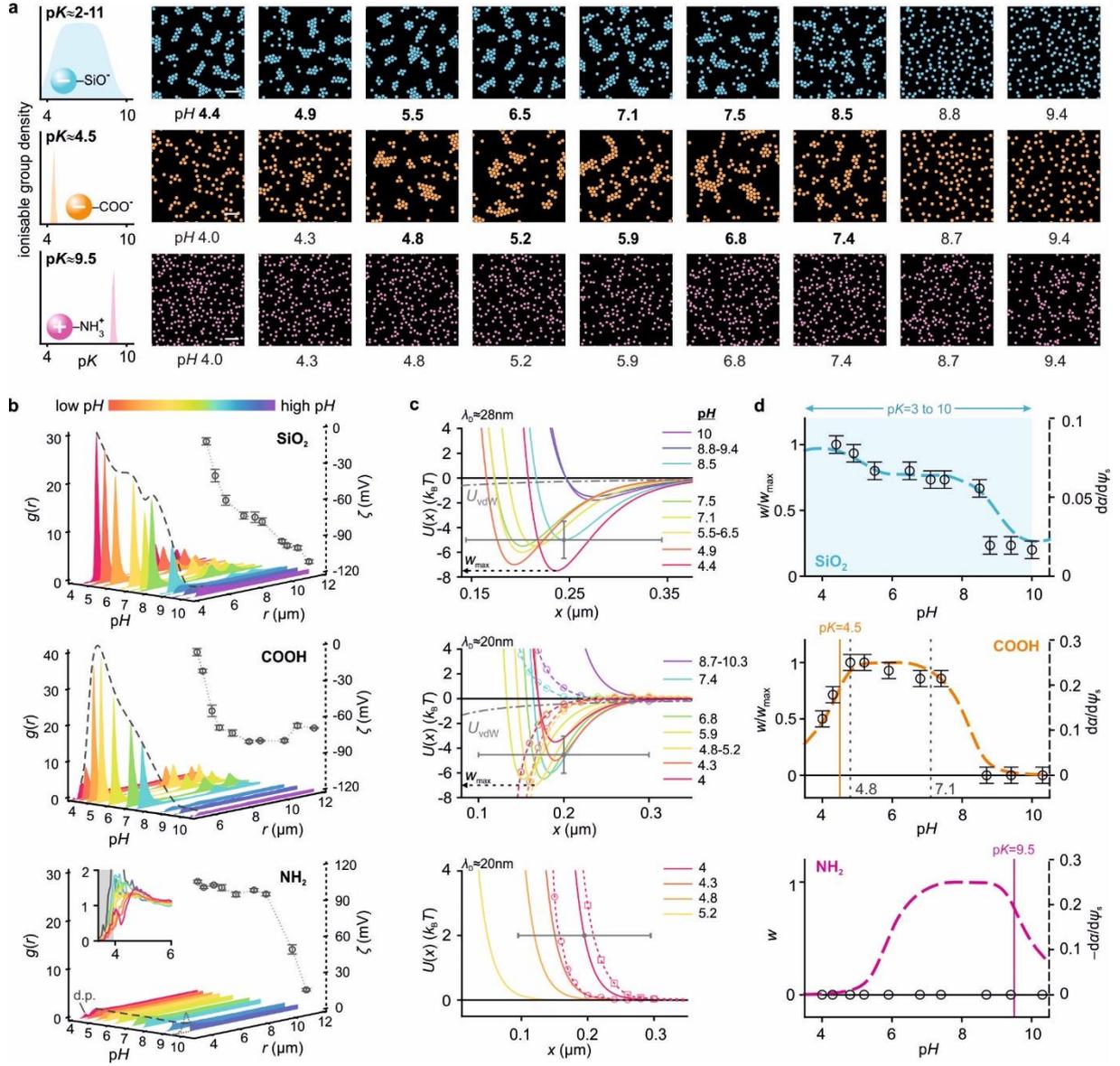

**Fig. 2. Tuning the interparticle interaction using the pH in solution.** (**a**) Snapshots of colloidal suspension structure as a function of $pH$ for silica particles (top), carboxylic acid particles (middle), and aminated silica particles (bottom), holding salt concentration constant in each data series: $c_0 =0.12$ (SiO$_2$), 0.25 (COOH), and 0.25 mM (NH$_2$) (scale bar: 20 μm). $pH$ at which hcp cluster formation is observed denoted in boldface. (**b**) Radial probability density functions $g(r)$ (curves) and measured $\zeta$ potentials (symbols) as a function of $pH$ for all three particle types. The increasing height of the "dimer-peak" (d.p.) with increasing $pH$ for positive particles is indicative of increased sticking of particles due to discharging of the basic ionisable groups (bottom panel). (**c**) $U(x)$ profiles inferred from measured $g(r)$ profiles (solid lines; see Supplementary Figs. S17-19 and Tables S4-8), theoretical $U_{tot}(x)$ curves (dashed lines with symbols) calculated as described in Ref.[27] and supplementary section S2.1 and 2.2 (see Supplementary Table S1 for parameter values), and $U_{vdW}$ contribution for comparison (dashed-dotted grey line). $w_{max}$ denotes the maximum inferred depth of the experimentally observed minimum in each case. (**d**) Plots comparing trends as a function of $pH$ in the experimentally inferred quantity $w/w_{max}$ (symbols) with that of $\frac{d\alpha}{d\psi_s}$ determined using Eq. 5 (dashed curves). We model silica using 7 different $pK$ values between 3 and 9 (top), whereas for COOH (middle) and NH$_2$ particles (bottom) we use single $pK$ values of 4.5 and 9.5 respectively (see supplementary section S2.5). For COOH particles (middle panel), the contribution from the interfacial free energy is maximal in the range $pK + 0.3 \lesssim pH \lesssim pK + 2.5$ (dotted vertical lines, see main text). For positively charged particles, the model does not envisage an interparticle attraction, implying $w = 0$ even when $-\frac{d\alpha}{d\psi_s}$ is large, in agreement with measurements (bottom). See Supplementary Fig. S22 for all raw image data.

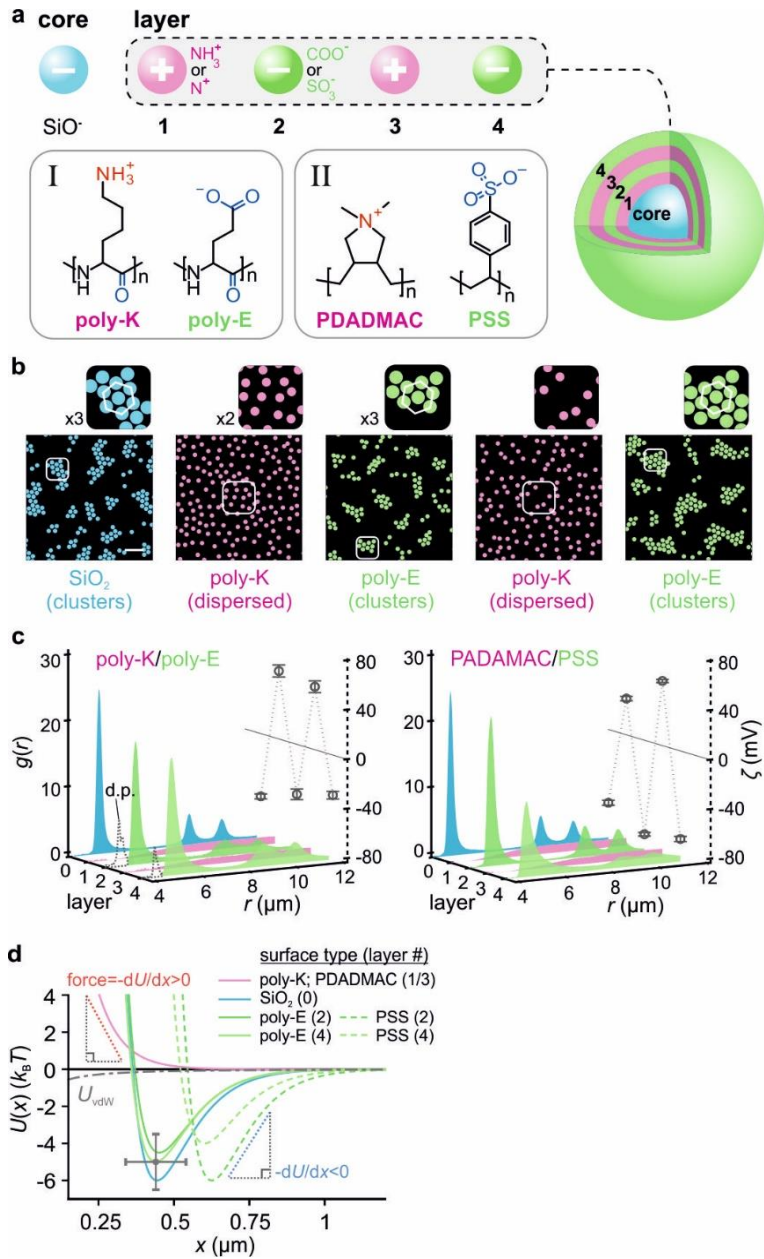

**Fig. 3. The sign of the long-range force depends on the sign of charge of the interacting particles.**
(**a**) Schematic depiction of a sequentially coated silica particle displaying 4 sequential layers with alternating sign of coating charge (top). Chemical structures of charged polypeptides (polylysine: poly-K, and polyglutamic acid: poly-E) and polyelectrolytes (poly-diallyldimethylammonium chloride: PDADMAC, and polystyrene sulfonate: PSS) used for layer-by-layer coating of silica particles (bottom). (**b**) Images of particle suspension structure following each coating procedure using charged polypeptides (scale bar: 20 μm). See Supplementary Fig. S20 and S22 for polyelectrolyte data, including PEI/PSS coatings, and raw images. (**c**) Radial probability density functions $g(r)$ and measured $\zeta$ potentials as a function of coating or layer number for polypeptide (left) and polyelectrolyte coatings (right). The alternating sign of measured $\zeta$ potential values for each new coating layer confirms the change in sign of particle charge. (**d**) $U(x)$ profiles inferred from measured $g(r)$ profiles for particles with positively charged (pink curve) and negatively charged coatings (blue and green curves) with layer numbers in parentheses (see Supplementary Fig. S20 and Tables S9-S11). At long range ($x \gtrsim 5\kappa^{-1}$), negatively charged particles ($\sigma < 0$) display an attraction which implies that the interparticle force is attractive, i.e., $-\frac{dU(x)}{dx} < 0$, whereas the converse is true for positively charged particles ($\sigma > 0$) where we infer $-\frac{dU(x)}{dx} > 0$ for all separations which agrees qualitatively with DLVO theory. All experiments performed at $c_0 = 15$ μM.

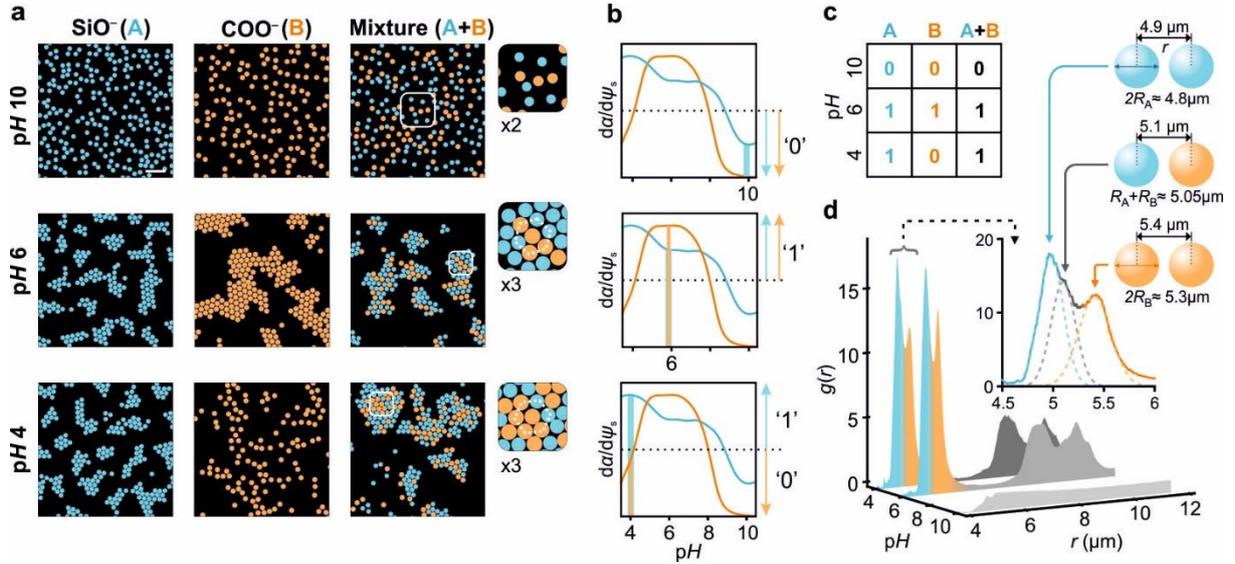

**Fig. 4. Cluster formation in chemically dissimilar particles**. (**a**) Structure of suspensions of pure SiO$_2$ particles (species A - blue, left), COOH particles (species B - orange, middle) and equimolar mixtures of the two species (right) at p$H$ 4, 6 and 10. hcp clusters containing both particle species form even under conditions where one of the particle species does not display significant self-attraction (p$H$ 4 for this system, bottom panels). (**b**) Plots of normalized $\frac{d\alpha}{d\psi_s}$ vs. p$H$ for SiO$_2$ (blue curve) and COOH (orange curve) from Fig. 2, with a dotted black line indicating a threshold level above which we regard $\frac{d\alpha}{d\psi_s}$ to be high, taking a binarized value of 1, which is accompanied by the formation of strong and stable same-species clusters as shown in Fig. 2. For low values of $\frac{d\alpha}{d\psi_s}$ we expect no cluster formation and assign this state and cluster formation outcome a value of 0. (**c**) Truth table for the outcome of cluster formation in mixtures (column 3) depending on the binarized values of $\frac{d\alpha}{d\psi_s}$ for the individual pure species (columns 1 and 2). A value of 1 indicates the formation of mixed clusters containing both particle species, and 0 reflects no cluster formation. (**d**) Radial probability distributions $g(r)$ for the 3 cases in (**a**), displaying clear peak signatures of all three expected nearest neighbour distances in the cross-species hcp clusters, i.e., $r_{AA} > 2R_A$, $r_{AB} > R_A + R_B$, $r_{BB} > 2R_B$ (inset).

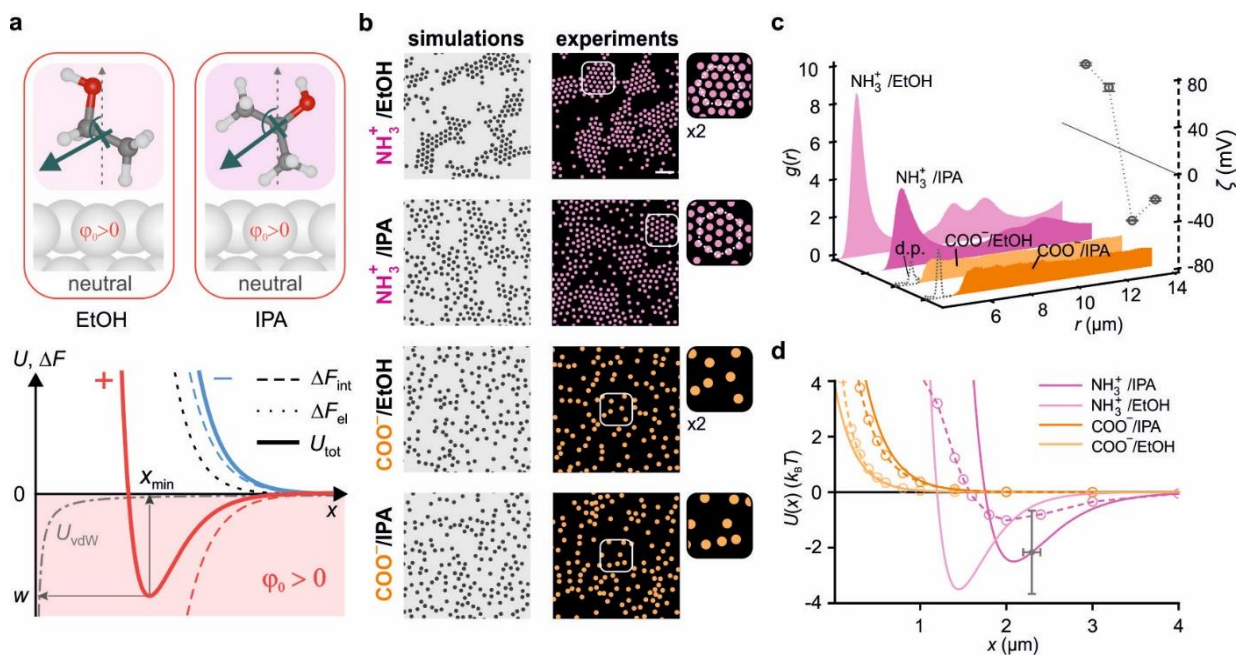

**Fig.5. Negative self-attraction in water switches to positive self-attraction in alcohols.** (**a**) MD simulations of ethanol and isopropanol at a neutral oxygen atom surface show that these molecules orient on average with their methyl groups pointing slightly towards the interface and their electronegative O-atoms pointing slightly towards the bulk, giving rise to an interfacial or solvation potential $\varphi_0 > 0$ (top). Unlike water, the average interfacial molecular dipole moment (green arrow) now points towards the surface. Fig. S8b presents the variation in interfacial potential, $\varphi_{\text{int}}$, and the corresponding interfacial free energy, $f_{\text{int}}$, as a function of surface charge density, $\sigma$, from MD simulations of isopropanol at an interface. Bottom: various contributions to the total free energy of interaction for the interaction of positively (red lines) and negatively charged (blue lines) particles: electrostatic repulsion, $\Delta F_{\text{el}}$ (black dotted line and applies to both signs of charge), interfacial solvation contribution, $\Delta F_{\text{int}}$ (dashed line, different for positive and negative), total interaction energy, $U_{\text{tot}} = \Delta F_{\text{el}} + \Delta F_{\text{int}}$ (solid line), and the small vdW contribution, $U_{\text{vdW}}$ (dashed-dotted line, applies to both types of particle). When $\varphi_0 > 0$, positively charged particles in solution are expected to display a minimum in $U_{\text{tot}}$ at an interparticle separation $x_{\text{min}}$ (solid red line), whereas negatively charged particles are expected to repel monotonically, displaying no minimum (solid blue line). (**b**) Experiments (right panels) show that positively charged aminated silica particles (pink) form hcp clusters, whereas negatively charged carboxylated particles (orange) do not form clusters in alcohols (scale bar: 20 μm, see Supplementary Fig. S22 for raw images). BD simulations with appropriate input pair interaction potentials recover the experimentally observed suspension structure (left panels). (**c**) Measured radial probability density functions, $g(r)$, for positively charged particles (shades of pink) and negatively charged particles (shades of orange) in alcohols, including measured ζ potentials (circular symbols) for each case. Experiments correspond to measured values of $c_0 \approx 0.4$ μM (see Supplementary Table S12 for details). (**d**) Pair interaction potentials of the form $U(x) = A\exp(-\kappa_1 x) + B\exp(-\kappa_2 x)$ inferred for the experimental data from BD simulations (solid lines) for negative particles (orange) and positive particles (pink) (see Supplementary Fig. S21 and Table S12 for details). Calculated $U_{\text{tot}}(x)$ curves as described in Ref. 27 and supplementary section S2.1 and 2.2 (dashed lines with symbols; see Supplementary Table S1 for parameter values).

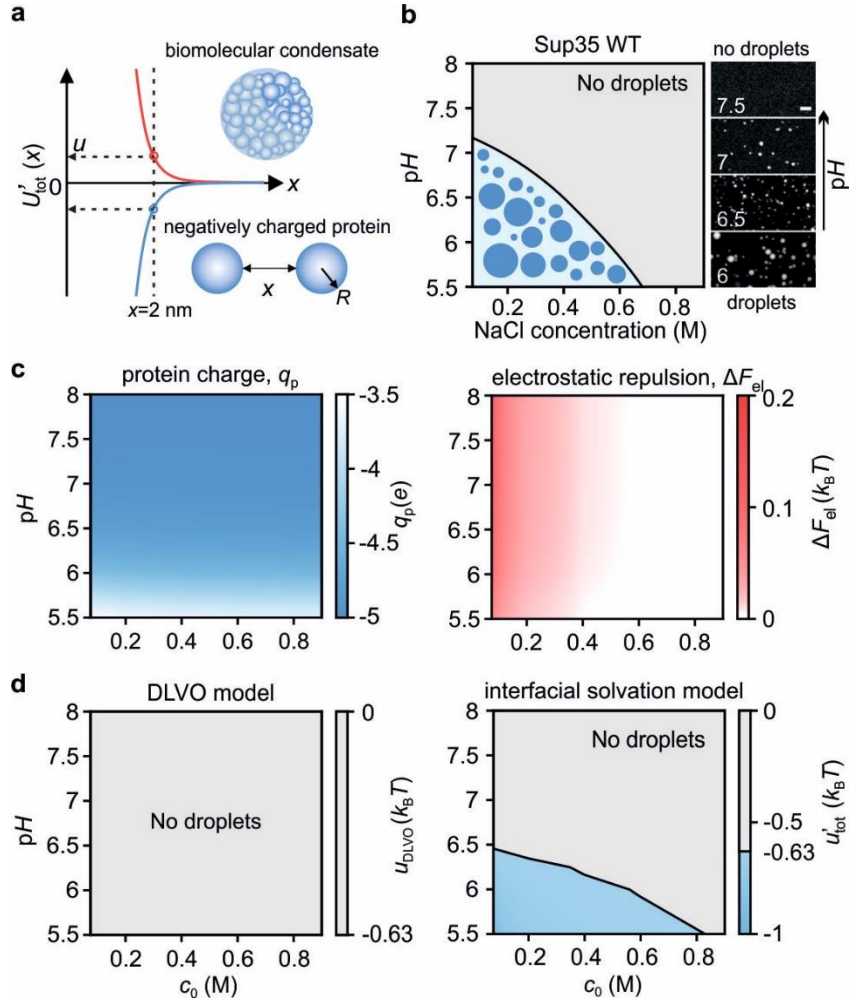

**Fig. 6. The charge dependent solvation force can drive the formation of biomolecular condensates.** (**a**) Schematic depiction of a biomolecular condensate formed from protein molecules carrying net negative electrical charge (light blue spheres). A model negatively charged protein is treated as a sphere of radius $R = 2.5$ nm carrying a charge equivalent to a net surplus of 5 glutamic acid groups. Blue and red lines present two possible, qualitatively different interaction potentials $U(x)$ for a pair of interacting molecules and we denote $u$ as the value of interaction energy at an intersurface separation of $x = 2$ nm in both scenarios (see supplementary section S2.7 for details). (**b**) Calculated $pH$ and salt concentration dependence of the phase separation of yeast wild-type protein Sup35 (left). The region shaded blue depicts conditions that support the formation of droplet condensates. Fluorescence images of condensate formation at a salt concentration of ≈0.1 M depict marked reduction and disappearance of droplets at $pH > 6.5$. Experimental data reproduced with permission from Ref. 47. (**c**) Calculated molecular charge shows that the maximum value of $q_p = -5\,e$ is attained at $pH > 6.5$ but that the charge of the molecule remains substantially negative over a broader range of $pH$ (left). Calculation of the electrostatic interaction energy $\Delta F_{el}(x = 2\,\text{nm})$ suggests a net repulsive interaction between two molecules for the relevant $pH$ and $c_0$ range (right). (**d**) Applying a pairwise interaction energy cut-off of $u \lesssim u_b = -0.63\,k_B T$ suggests no phase separation from a DLVO perspective (left). Inclusion of the solvation free energy contribution yields a coexistence curve which indicates $pH$ and salt concentration dependence of droplet formation similar to the experimental observations (right) (see supplementary section S2.7 for details).

27